\documentclass[11pt,a4paper]{article}
\usepackage{jheppub}

\newcommand{\as}{\alpha_s}
\newcommand{\mupi}{\mu_\pi^2}

\newcommand{\ep}{\epsilon }
\newcommand{\uh}{{\hat u} }

\long\def\symbolfootnote[#1]#2{\begingroup%
\def\thefootnote{\fnsymbol{footnote}}\footnote[#1]{#2}\endgroup}

\def \be{\begin{equation}}
\def \ee{\end{equation}}
\newcommand{\bea}{\begin{eqnarray}}
\newcommand{\eea}{\end{eqnarray}}
\def \nn{\nonumber}

\numberwithin{equation}{section}
\allowdisplaybreaks[4]

\title{\boldmath Perturbative corrections to power  suppressed
effects in $\bar B\to X_u\ell\nu$ }
\author[a]{Bernat Capdevila,}
\author[a]{Paolo Gambino,}
\author[b]{and Soumitra Nandi\,}

\affiliation[a]{Universit\`a di Torino, Dip.\ di  Fisica \& INFN, Torino\\ Via Giuria 1, Torino, I-10125, Italy}
\affiliation[b]{Dept.\ of Physics,
Indian Inst.\ of Technology Guwahati\\ 781 039, India}

\abstract{
  We compute the $O(\alpha_s)$ corrections to the Wilson coefficients of the dimension five operators
  in inclusive semileptonic $B$ decays in the limit of a massless final quark.
  Our calculation agrees with reparametrization invariance and with  previous results for the total width and improves the constraints on  the shape functions that enter those decays.}

\begin{document}
\maketitle

\section{Introduction}
Despite a significant experimental effort at the $B$ factories, the current status of the determination of the  CKM matrix element $V_{ub}$ is far from  satisfactory.
The magnitude of $V_{ub}$ is determined from  semileptonic $B$ decays without charm
and  in the inclusive case stringent phase-space cuts must be employed to suppress the dominant $B\to X_c \ell \nu$ background.
The modern description of these inclusive  decays is based on a non-local Operator Product Expansion (OPE) \cite{Neubert:1993ch,Bigi:1993ex},
where nonperturbative shape functions (SFs) play the role of parton distribution functions
of the $b$ quark inside the $B$ meson.
Among the theoretical frameworks that incorporate this formalism, BLNP \cite{Lange:2005yw},
GGOU \cite{Gambino:2007rp}, and  DGE
\cite{Andersen:2005mj}
are currently employed by the Heavy Flavour Averaging Group (HFLAV)  \cite{Amhis:2019ckw}. The latest average values of $|V_{ub}|$ in these three
frameworks,
\[
|V_{ub}|^{\rm BLNP}\!\!=4.44(26) \times 10^{-3},\quad
|V_{ub}|^{\rm GGOU}\!\!=4.32(18) \times 10^{-3},\quad
|V_{ub}|^{\rm DGE}\!\!=3.99(14) \times 10^{-3},
\]
do not agree well with each other. Moreover, the values obtained from different experimental
analyses are not always compatible within their stated theoretical and experimental
uncertainties.
The latest endpoint analysis by BaBar \cite{TheBABAR:2016lja}, in particular, shows a strong dependence on the model used to simulate the signal and leads to sharply different results
in BLNP and GGOU. This  is the most precise analysis to date; in GGOU and DGE it favours a lower $|V_{ub}|$  and it is therefore in better
agreement  with
\be
|V_{ub}|_{av}^{B\to \pi\ell\nu}= 3.70(16) \times 10^{-3}, \label{eq:1}
\ee
the value extracted from $B\to \pi \ell\nu$ data together with lattice QCD determinations of the relevant form factor \cite{Amhis:2019ckw}.
It is also worth mentioning that a preliminary tagged analysis based on the full Belle data set
\cite{Cao:2021xqf}
indicates a better  agreement both among theoretical frameworks and with Eq.~(\ref{eq:1}).

The large statistics available at Belle II should help clarify  the matter in various ways,  see \cite{Gambino:2020jvv}. In particular, it should be possible to calibrate and validate the different frameworks directly on data, especially on differential distributions which are sensitive to the SFs. The SIMBA \cite{Ligeti:2008ac, Bernlochner:2020jlt} and  NNVub \cite{Gambino:2016fdy} methods both aim at a model-independent parametrisation of the relevant SFs and are well posed to analyse the future Belle II data in an efficient way.

In view of these interesting prospects, various improvements are necessary on the theoretical side, among which the inclusion of   $O(\alpha_s^2)$ corrections not enhanced by $\beta_0$ \cite{Brucherseifer:2013cu} and of
$O(\alpha_s/m_b^2)$
effects that modify the OPE constraints on the SFs.
The latter corrections have been computed at the level of form factors (and therefore of the triple differential distribution) for the inclusive decays to charm \cite{Alberti:2012dn, Alberti:2013kxa}, see also \cite{Becher:2007tk, Mannel:2015jka}, but due to the intricate
interplay of soft and collinear singularities the limit of $m_c\to 0$ is far from trivial,
especially since in the case at hand the infrared singularities are power-like.
One possibility is to repeat  the calculation setting  $m_c=0$  from the start, but
we will show instead that the $m_c\to 0$ limit can be taken in a conceptually simple
manner, reproducing the expected pattern of collinear and soft-collinear singularities, as well
as a few  existing results.

Our method consists in systematically disentangling  all singularities that emerge in the $m_c\to 0$ limit  at the level of the form factors $W_i$; since the phase space integrals
of the form factors
are infrared safe, one can reorganise them in such a way to remove  the mass singularities completely. In this way we obtain  analytic results for both $O(\as \mu_\pi^2/m_b^2)$ and
$O(\as \mu_G^2/m_b^2)$ corrections to the form factors and therefore to the triple differential distribution.
Our results for the $O(\as \mu_\pi^2/m_b^2)$ corrections satisfy the reparametrization invariance relations obtained in \cite{Manohar:2010sf}, while the $O(\as \mu_G^2/m_b^2)$ corrections reproduce the shift in the total width computed at $m_c=0$ in Ref.~\cite{Mannel:2015jka}.
We also use our results to compute the $O(\as)$ corrections to the $q_0$-moments of the individual form factors,  which place crucial constraints on the SFs.

The outline of this paper is as follows. In section 2 we introduce our notation and review the
known $O(\as)$  corrections to the triple differential rate in the charmed case.
Section 3 gives an elementary illustration of  our method, taking the limit $m_c\to 0$ of the $O(\as)$  corrections and recovering the known results.
In section 4 we apply the method to the $O(\as\Lambda^2/m_b^2)$ corrections, with all analytic results given in the Appendix.
In section 5 we check that our results for the $O(\as \mupi/m_b^2)$ satisfy the reparametrization invariance relations. Section 6 is devoted to a few  applications: we compute the total
decay rate, the $q^2$ spectrum,  and the first moments of the form factors. Finally,
section 7 summarises our findings.

\section{\boldmath  Notation and ${ O(\as)}$ corrections}
We will consider the decay of a $B$ meson
of four-momentum $p_B= M_B v$ into a lepton pair with momentum $q$ and
a hadronic final state with momentum $p'=p_B-q$. Let us first assume that the
hadronic final state contains a charm quark with mass $m_c$
and express the $b$-quark decay kinematics in terms of the dimensionless quantities
\be
\rho= \frac{m_c^2}{m_b^2}, \qquad\quad \hat u= \frac{(p-q)^2 - m_c^2}{m_b^2} ,\quad\qquad \hat q^2= \frac{q^2}{m_b^2},
\ee
where $p= m_b v$ is the momentum of the $b$ quark and the physical range is given by
\be
0\le \hat u \le \hat u_+ =(1-  \sqrt{\hat{q}^2})^2 -\rho \qquad  {\rm and} \qquad
0\le \hat q^2 \le (1-\sqrt{\rho})^2.
\label{physrange}
\ee
We will also employ the energy of the hadronic system normalized to the $b$ mass
\be
E=  \frac12 (1+\rho +\hat u -\hat q^2).
\ee
The case of tree-level kinematics corresponds to $\hat u=0$; we indicate the corresponding   energy of the hadronic final state as
\be
E_0= \frac12 (1+\rho -\hat q^2).
\ee
 The  normalized total leptonic energy is
\be
\hat q_0 =1-E \quad\quad {\rm from\ which\ follows } \quad\quad \hat u = 2\,(1-E_0-\hat q_0).
\ee
We also introduce a threshold factor
\be
\lambda=4\,(\hat q_0^2-\hat q^2)=
4\,(E^2-\rho-\hat u). 
\ee
In the case of tree-level kinematics, the threshold factor becomes $\lambda_0=4(E_0^2-\rho)$.
It is convenient to introduce a short-hand notation for the square root of $\lambda$:
\be
t= \frac{\sqrt{\lambda}}{2\, E}, \qquad\quad t_0= \frac{\sqrt{\lambda_0}}{2\, E_0}. \label{tdef}
\ee
The differential $B\to X \ell\nu$ decay rate is proportional to  the product of a leptonic
and a hadronic rank-2 tensors, where the hadronic tensor $W^{\mu\nu}$ describes all the QCD dynamics in the decay.
It is customary to decompose  $W^{\mu\nu}$ into form factors,
\be
m_b \,W^{\mu\nu}(p_B,q)=-W_1 \, g^{\mu\nu}+W_2 \,v^\mu v^\nu +i W_3 \, \ep^{\mu\nu\rho\sigma }
v_\rho \hat q_\sigma + W_4 \hat q^\mu \hat q^\nu +W_5 \left(v^\mu \hat q^\nu\!+\! v^\nu \hat q^\mu\right),
\ee
where $\hat q^\mu=q^\mu/m_b$, $v^\mu$ is the four-velocity of the $B$ meson, and
 the  $W_i$ are functions of $\hat q^2$ and $\hat q_0$, or equivalently of
$\hat q^2$ and $\hat u$.

 In the limit of massless leptons only $W_{1,2,3}$ contribute to the decay rate and one has
\bea
\frac{d \Gamma}{\, d\hat E_\ell \,d\hat q^2 \, d\hat u }&=&
\frac{G_F^2 m_b^5 |V_{cb}|^2}{16\pi^3}
\theta(\hat u_+ -\hat u)\theta(\hat E_\ell) \theta(\hat q^2)\times \\ &&
\times\left\{
\hat q^2\, W_1 -\left[2 \hat E_\ell^2-2\hat E_\ell\hat q_0 +\frac{\hat q^2}2 \right] W_2 +\hat q^2 (2\hat E_\ell -\hat q_0)\,W_3\right\} ,\nn
\label{rate}
\eea
where $\hat u_+$, defined in (\ref{physrange}),
represents the kinematic boundary on $\hat u $, and $\hat E_\ell=E_\ell/m_b$ is the normalized charged lepton energy.
Thanks to the OPE, the structure functions can be expanded in series of $\alpha_s$ and $\Lambda_{\rm QCD}/m_b$. There is no term linear in $\Lambda_{\rm QCD}/m_b$ and therefore
\be
W_i =W_i^{(0)} + \frac{\mu_\pi^2}{2m_b^2} W_i^{(\pi,0)}+\frac{\mu_G^2}{2m_b^2} W_i^{(G,0)}+
  \frac{\as}{\pi}\left[C_F  W_i^{(1)} +C_F
  \frac{\mu_\pi^2}{2m_b^2} W_i^{(\pi,1)}+\frac{\mu_G^2}{2m_b^2} W_i^{(G,1)}\right]
\ee
where we have neglected terms of higher order in the expansion parameters.
$\mu_\pi^2$ and $\mu_G^2$ are the $B$-meson matrix elements of the only gauge-invariant
dimension 5 operators that can be formed from the $b$ quark and gluon fields  \cite{Bigi:1992su,Blok:1993va}.
In the Standard Model the leading order coefficients are given by
\be
W_i^{(0)} = w_i^{(0)} \,\delta(\hat u);  \qquad  \qquad w_1^{(0)} = 2 E_0, \qquad w_2^{(0)} = 4, \qquad w_3^{(0)} =2.
\ee
The tree-level nonperturbative coefficients  $W_i^{(\pi,0)}$ and $W_i^{(G,0)}$
\cite{Blok:1993va}  are given in compact form in \cite{Alberti:2012dn, Alberti:2013kxa}.
The leading perturbative corrections to the free quark decay have been computed in \cite{Aquila} and refs.\ therein. They read
\be
W^{(1)}_i=w_i^{(0)} \left\{ S_i \,\delta(\hat u) -2 \left(1 -  E_0 I_1\right)
 \left[\frac1{\hat u}\right]_+   +\frac{\theta(\hat u)}{(\rho + \hat u)}   \right\} + R_i  \,\theta(\hat u)
\label{NLO1},
\ee
where $S_i= S+\Delta_i$ and
\bea
S&=&2E_0 \left(I_{2,0}-I_{4,0}\right) 
-1 -\frac{1-\rho-6\hat q^2}{4\hat q^2} \ln \rho 
-\frac{(1 - \rho)^2 - 6\,  \hat q^2 (1 + \rho) + 5 (\hat q^2)^2 }{4\hat q^2}  I_{1,0}\,;\nn\\
 \Delta_1\!&=&\!\!-\frac{\rho}{E_0} I_{1,0}; \qquad \Delta_2= \frac{1-\rho}{4\hat q^2} \ln \rho+\left(\frac{(1-\rho)^2}{4\hat q^2}-\frac{1+\rho}{4}\right)I_{1,0}; \qquad \Delta_3=0, \label{eqS}
\eea
and the functions $R_i$ are given in Eqs.~(2.32-2.34) of Ref.~\cite{Aquila}.\footnote{The
variables $\hat \omega $, $\lambda_b$, and $\tau$ of Ref.~\cite{Aquila} correspond to $-2 E_0$, $\lambda$ and $(1-t)/(1+t)$, respectively.}
 The integrals
$I_1$, $I_{1,0}$, $I_{2,0}$, and $I_{4,0}$ are given  in Eqs.~(A.6-8) of \cite{Alberti:2012dn}
and the plus distribution is defined by
its action on a generic test function $f(\hat u)$:\footnote{
Ref.~\cite{Aquila} uses $\hat u_+$ as upper limit, and the two definitions can be easily
 related, see \cite{Alberti:2012dn}.}
\be
\int f(\hat u) \left[\frac1{\hat u}\right]_{+}d\hat u =
\int_0^{1} \frac{f(\hat u)-f(0)}{\hat u}\,d\hat u.
\label{plus1}
\ee

\section{The massless limit}
We now take the limit $m_c\to 0$, {\it  i.e.}\ $\rho\to 0$, of the $O(\as)$ corrections to the form factors, $W_i^{(1)}$. Of course, collinear divergences
emerge in this way, leading to $\ln \rho$ and $ \ln^2 \rho$ in  $W_i^{(1)}$, which however are
compensated upon integration  over $\hat u$, as collinear logs  arise from the phase space integration as well. As the phase space integrals of $W_i$  are infrared safe, one can therefore reorganise the
expressions for $W_i$ in order to remove completely the mass singularities. In practice it is
sufficient to  consider the integral
\be
\int f(\hat u) \,W_i^{(1)}(\hat u,\hat q^2) \,d \hat u,
\ee
where $f(\hat u) $ is a generic test function.

Let us first consider the limit for $\rho\to 0$ of the coefficient of the $\delta(\hat u)$, the function $S$ given in (\ref{eqS}). The integrals $I_{k,0}$ admit the simple expansions
\bea
I_{1,0}&=& \frac{2 \ln (1-\hat q^2) -\ln \rho}{1-\hat q^2} + O(\rho)\label{limitI10}\\
I_{2,0}&=&\frac{ {\rm  Li}_2( \hat q^2)-\frac{\pi^2}6}{ 1 - \hat q^2} + O(\rho)\\
I_{4,0}&=&\frac{ 2 \ln^2 (1-\hat q^2) +2{\rm  Li}_2( \hat q^2)-\frac12 \ln^2 \rho}{ 1 - \hat q^2} + O(\rho)
\eea
and we therefore have
\be
S=\frac{\ln \rho}4 +\frac{\ln^2 \rho}2 - \frac{\pi^2}6- {\rm Li}_2(\hat q^2)-2 \ln^2 (1-\hat q^2)-1-
\frac{1-5\hat q^2}{2\hat q^2} \ln (1-\hat q^2)+O(\rho)
\ee
We now consider the real emission contributions given by $R_i$. Their structure is
\be
R_i= \frac{ r_i^{(1)}\hat u +  r_i^{(2)}\rho}{(\hat u+\rho)^2}+\frac{s_i}{\hat u+\rho}+t_i
\label{Ri}
\ee
where $r_i, s_i, t_i$ are functions of $\hat q^2$ and $\hat u$ that are regular in the limit $\hat u, \rho \to 0$. Clearly, the collinear singularities at $\hat u=0$ are regulated  by $\rho$. To expose them, let us start with the second term in (\ref{Ri}) and observe that for  a  test function $f(\hat u)$
\[
\int_0^1  f(\hat u)\, \frac1{\hat u + \rho}  \,d\hat u=\int_0^1  \frac{f(\hat u)-f(0)+f(0)}{\hat u + \rho}  \,d\hat u= \int_0^1 f(\hat u) \left(\left[\frac1{\hat u}\right]_+\! -\ln \rho\, \delta(\hat u) \right)d\hat u+O(\rho)
\]
and therefore in the second term of (\ref{Ri}) and in the last term in the universal part of (\ref{NLO1}) we can safely make the replacement
\be
\frac1{\hat u+\rho}\to \left[\frac1{\hat u}\right]_+ -\ln \rho\, \delta(\hat u) \label{d1rule}
\ee
and take the limit of $s_i$ for $\rho\to 0$. This extracts one of the singularities we were looking for. Let us now turn to the first term in (\ref{Ri}). In the limit $\rho\to 0$ the coefficient of $r_i^{(2)}$ in (\ref{Ri}) is proportional to $\delta(\hat u)$.   Taking into account that
\be
\int_0^1 \frac{\rho}{(\hat u + \rho)^2}\ d\hat u=  1+O(\rho),
\ee
we can therefore use the  replacement
\be
 \frac{\rho}{(\hat u + \rho)^2} \to \delta(\hat u),
\ee
and take the limit of $r_i^{(2)}$ for $\rho\to 0$.
A linear combination of the two above replacement rules deals with
the coefficient of $r_i^{(1)}$, $\hat u/(\hat u +\rho)^2$.

Let us now consider the plus distribution
in (\ref{NLO1}), and in particular  the part involving $I_1$. Here the singularity is hidden in the integral $I_1$ and in its $\hat u\to 0$ limit. They are given by
 \be
I_1=I_1(\hat q^2, \hat u)=   \frac{\ln\frac{1+t}{1-t}}{\sqrt{\lambda} } ,\qquad\qquad
I_{1,0}= I_{1}(\hat q^2, 0)=\frac{\ln\frac{1+t_0}{1-t_0}}{\sqrt{\lambda_0} },
\label{I1}
 \ee
where $t$ and $t_0$ have been introduced in (\ref{tdef}).
Let us first focus on
\be
\frac{I_1 -I_{1,0}}{\hat u} \label{pippo}
\ee
 which is  a function of $\hat u, \rho$, and $\hat q^2$ and is non-analytic at $\hat u=\rho=0$.  Indeed, expanding (\ref{pippo}) for $w=1-\hat{q}^2\gg \hat u, \rho$
we find that its leading singularity is
\be
 \frac{I_1 -I_{1,0}}{\hat u}\Big|_{sing}= \frac{\ln \frac{\rho}{\hat u + \rho}}{w\hat u} . \label{pippo2}
\ee
The difference of (\ref{pippo}) and (\ref{pippo2}) is however regular in the limit $\rho\to 0$, and  we can  split (\ref{pippo}) into a singular and a regular piece,
\be
\frac{I_1 -I_{1,0}}{\hat u} = \frac1{w\hat u}\ln \frac{\rho}{\hat u + \rho} + B(\hat q^2 , \hat u) + O( \rho).\label{pippo3}
\ee
Denoting by ${\cal I}_1$ the limit of $I_1$ for   $\rho\to 0$, the function $B$ is given by
\be
B(\hat q^2, \hat u)=\frac{ \ln (\hat u/w^2)+w  {\cal I}_1}{w \hat u}\simeq
 \frac{w-2}{w^3} \ln \frac{\hat u}{w^2} + \frac{2(w-1)}{w^3}+O(\hat u)\label{eq:Bexp}
\ee
which has only a logarithmic (integrable) singularity in $\hat u$ and can be considered regular for our purposes.
We can now use the definition of the plus distribution with a test function $f(\hat u)$ and
reorganize the integral  as follows:
\bea
&&\int  f(\hat u) \  I_1\Big[ \frac{1}{\hat u}\Big]_+ d\hat u=\int_0^1  \frac{f(\hat u) I_1(\hat u) - f(0) I_{1,0}}{\hat u} d\hat u \label{eq:1.21}\\
&&= f(0) \int_0^1 \frac{I_1 -I_{1,0}}{\hat u} d \hat u + \int_0^1 \frac{(f(\hat u ) - f(0)) (I_1 -I_{1,0})}{\hat u} d \hat u+ I_{1,0} \int_0^1 \frac{f(\hat u ) - f(0)}{\hat u} d\hat u.\nonumber
\eea
Keeping in mind that we can drop all $O(\rho)$ terms, the first term in the last line
is the sum of the integrals of the first two terms on the rhs of (\ref{pippo3}). We can  simplify
the second term by using (\ref{pippo3}) again, and obtain several terms, among which a logarithmic plus distribution,
which signals the appearance of the soft-collinear divergence.
Finally, in the last term we can use the $\rho\to 0$ expansion of $I_{1,0}$ given in (\ref{limitI10}). The result is\footnote{We do not display a $\theta(1-\hat u)$ that arises from the
above calculation, as it would be irrelevant for any practical application.}
\be
\int  f(\hat u) \  I_1\Big[ \frac{1}{\hat u}\Big]_+ d\hat u= \int f(\hat u) \Big[ a\, \delta(\hat u) + b\, \Big[\frac{\ln\hat u}{\hat u}\Big]_+ +c\, \Big[\frac{1}{\hat u}\Big]_+ +d \,\theta(\hat u)\Big]d\hat u
\ee
with
\be
a= -\frac{\frac{\pi^2}3+\ln^2\rho}{2w}, \qquad b=- \frac1{w}, \qquad
c= \frac{2\ln w}{w}, \qquad
d= B(\hat q^2, \hat u).
\ee
 Notice that the integrals of $ B( \hat q^2, \hat u)$ in the first and second term of the second line of (\ref{eq:1.21}) cancel each other.

We are now in the position to take the limit for $\rho\to 0$ of the whole $W_i^{(1)}$.
Collecting all terms 
we verify  that the
mass singularities cancel completely and  obtain, with  $w=1-\hat q^2$,
\be
W^{(1)}_i=w_i^{(0)} \left\{  {\cal S}_i \,\delta(\hat u) -
 \left[\frac{\ln \hat u}{\hat u}\right]_+ - \Big(\frac74 - 2\ln w\Big) \left[\frac{1}{\hat u}\right]_+ +
 w \, B(\hat q^2, \hat u)
\, \theta(\hat u)
   \right\} + {\cal R}_i^{(1)}  \,\theta(\hat u)
\label{NLO},
\ee
where 
\be
{\cal S}_i=-\frac54 - \frac{\pi^2}3 - {\rm Li}_2(1-w)-2 \ln^2 w - \frac{5w-4}{2(1-w)} \ln w+  \frac{\ln w}{2(1-w)}\, \delta_{i2}
\ee
and the functions ${\cal R}_i^{(1)}$ are given by
\bea
{\cal R}_1^{(1)}&=&\frac34 + \frac{\hat u (12-w-\hat u)}{2\tilde\lambda}+
\Big( w + \frac{\hat u}2 -
\frac{ \hat u(2\hat u +3 w)  }{\tilde \lambda} \Big){\cal I}_1
\\
{\cal R}_2^{(1)}&=&\frac{6\hat u(\hat u^2-(3-w)\hat u -12+13 w)}{\tilde\lambda^2}+\frac{\hat u -38+21 w}{\tilde\lambda} 
\nonumber\\
&&
-4\frac{\frac{w}2 \hat u^3+ (2w^2-6)\hat u^2+ (7-3w+\frac52 w^2)w\hat u+ w^3(w-4)  }{ \tilde\lambda^2}\, {\cal I}_1
\\
{\cal R}_3^{(1)}&=& \frac{3\hat u -8+5w}{\tilde\lambda} + 
\frac{ \hat u^2 - (6-w)\hat u +4w }{ \tilde\lambda} {\cal I}_1
\eea
with
\be
{\cal I}_1= \frac1{\sqrt{\tilde\lambda}} \ln \frac{\hat u + w +\sqrt{\tilde\lambda}}{\hat u + w -\sqrt{\tilde\lambda}}
\ee
and $\tilde \lambda= (\hat u+w)^2-4 \hat u$.
These results are in complete agreement with the calculation of
$W_i^{(1)}$ with $m_c=0$ performed in  Ref.~\cite{DeFazio:1999ptt}.

\section{\boldmath The  $ O(\as \Lambda^2/m_b^2)$ results}
The method employed in the previous section can be readily extended to take the $m_c\to 0$ limit of the $O(\as \Lambda^2/m_b^2)$ results obtained in Refs.~\cite{Alberti:2012dn,Alberti:2013kxa}. The main difference is that perturbative corrections to power suppressed effects induce power-like divergences, including  collinear  power divergences in the $m_c\to 0$.
On the other hand, the most complicated features  of these singularities are determined by the same integral $I_1$ that we have encountered in the previous section, as the calculations of the $O(\as)$ and $O(\as \Lambda^2/m_b^2)$ corrections are based on the same
building blocks (master integrals).
The divergences in the
corrections related to the kinetic operator and proportional to $\mu_\pi^2$ are  stronger than
in those proportional to $\mu_G^2$. It is therefore instructive to start reviewing the structure of the $O(\as \mu_\pi^2/m_b^2)$ contributions for finite charm mass:
\bea
\label{Wipi1}
W^{(\pi,1)}_i&=& w_i^{(0)} \frac{\lambda_0}3 \Big(S_i +3 (1-E_0 I_{1,0})\Big)\delta''(\hat u)
 +b_i \,\delta'(\hat u) + c_i \,\delta(\hat u)\\
&&+ \,d_i\left[\frac1{\hat u^3}\right]_++e_i \left[\frac1{\hat u^2
 }\right]_+  + f_i \left[\frac1{\hat u
 }\right]_+ + R_i^{(\pi)}  \,\theta(\hat u) \nn ,
\eea
where the generalized plus distributions are defined by
\be
\label{plusn}
\int  \left[\frac{\ln^n \uh}{\uh^m}\right]_+  f(\uh)\,d\uh =\int^{1}_0
\frac{\ln^n \uh}{\uh^m} \left[f(\uh)-\sum_{p=0}^{m-1} \frac{\uh^p}{p!} f^{(p)} (0)\right]d\uh
\ee
with $f^{(p)}(\uh)= \frac{d^p f(\uh)}{d \uh^p}$, and $d_i, e_i, f_i$ are  functions of $\hat q^2$ and $\hat u$  linear in $I_1$.
The remainder terms
 $R_i^{(\pi)} $ can be written as
\be
R_i^{(\pi)}=  \frac{ p_i^{(1)}\hat u +  p_i^{(2)}\rho}{(\hat u+\rho)^4}+ \frac{ q_i}{(\hat u+\rho)^3}+\frac{ r_i}{(\hat u+\rho)^2}+\frac{s_i}{\hat u+\rho}+t_i,
\label{Ri2}
\ee
where $p_i^{(j)},q_i, r_i, s_i, t_i$ are also functions of $\hat q^2$ and $\hat u$ that are regular in the limit $\hat u, \rho \to 0$.
Notice that    the expressions for the $R_i^{(\pi)}$
given in  \cite{Alberti:2012dn} have a different form, as they also contain powers of $\hat u$ in the denominators. This is because Ref.~\cite{Alberti:2012dn} reduces
the coefficients of the plus distributions by Taylor expanding them around $\hat u=0$, namely employs
 \be
 f(\hat u) \left[\frac1{\hat u^2}\right]_+=f(0)\left[\frac1{\hat u^2}\right]_+ + f'(0) \left[\frac1{\hat u}\right]_+ + \frac{f(\hat u)-f(0)-\hat u f'(0) }{\hat u^2} , \label{idplus}
 \ee
and similar identities which simplify the coefficients of the plus distributions. However, in Ref.~\cite{Alberti:2012dn} such identities have been applied for finite $\rho$. The non-analyticity of $I_1$ at $\rho=\hat u=0$ implies that the limit $\rho\to 0$ should be taken {\it before} simplifying the coefficients of the plus distributions. We have therefore used the results of the calculation \cite{Alberti:2012dn} before the final simplifications.

Working in the same way as we did  after (\ref{Ri}) and using the definition (\ref{plusn})
of the generalized plus distributions, we can isolate the divergences in $R_i^{(\pi)}$. For instance, let us consider
\be
K=\int_0^1 f(\hat u)\, \frac1{(\hat u + \rho)^2} \, d\hat u
\ee
where $f(\uh)$ is again a generic test function. Subtraction of the divergent parts leads to
\be
K= \int_0^1 \frac{f(\hat u)-f(0)- \hat u f'(0)}{(\hat u + \rho)^2} \, d\hat u+f(0) \int_0^1  \frac1{(\hat u + \rho)^2} \, d\hat u+ f'(0) \int_0^1 \frac{\hat u}{(\hat u + \rho)^2} \, d\hat u
\ee
where the last two integrals can be solved and expanded in $\rho$, while the first has no mass singularity and after setting $\rho=0$ corresponds to the action of $[1/(\hat u^2)]_+$ on $f(\hat u)$.  We therefore find the replacement rule
\be
\frac1{(\hat u + \rho)^2}\to \Big[\frac1{\hat u^2}\Big]_+ +\Big( \frac1{\rho}-1\Big)\, \delta(\hat u)+
(\ln\rho+1) \ \delta'(\hat u)
\ee
and proceeding in a similar way we also find
\bea \frac1{(\hat u + \rho)^3}&\to& \Big[\frac1{\hat u^3}\Big]_+ +\frac12 \Big( \frac1{\rho^2}-1\Big)\, \delta(\hat u)-
\Big(\frac1{2\rho}-1\Big) \ \delta'(\hat u) -\frac12\Big(\ln \rho+\frac32\Big)\,\delta''(\hat u)\\
\frac{\rho}{(\hat u + \rho)^4}&\to&  \frac1{3\rho^2} \delta(\hat u)-
\frac1{6\rho} \ \delta'(\hat u) +\frac1{6} \delta''(\hat u)
\eea
where the power divergences in $\rho$ have become  apparent. These rules together with (\ref{d1rule}) allow us to isolate the singularities of $R_i^{(\pi)}$ in the limit of vanishing  $\rho$.
Like in the case studied in the previous section, the coefficients of the plus distributions contain the integral $I_1$ and one has to disentangle the collinear singularities starting from the definition of the plus distributions.

As a preliminary step in that direction let us consider the action of a third-order plus-distribution on the product of $I_1$ and a generic  test-function $f(\uh)$. It can be rearranged in the following way
\begin{align}\label{1u3decomp}
\int f(\hat{u})I_1\left [\dfrac{1}{\hat{u}^3}\right]_+d\hat{u}=&\;f(0)\int_0^1 \dfrac{I_1-I_{1,0}-\hat{u}\,I_{1,1}-\frac12 \hat{u}^2\,I_{1,2}}{\hat{u}^3}\,d\hat{u}\nn\\
&+f^\prime(0)\int_0^1 \dfrac{I_1-I_{1,0}-\hat{u}\,I_{1,1}}{\hat{u}^2}\,d\hat{u}\\
&+\dfrac{f^{\prime\prime}(0)}{2}\int_0^1 \dfrac{I_1-I_{1,0}}{\hat{u}}\,d\hat{u}+I_{1,0}\int_0^1 f(\hat{u})\left[\dfrac{1}{\hat{u}^3}\right]_+ d\hat{u}\nonumber\\
&+\int_0^1 \dfrac{(f(\hat{u})-f(0)-\hat{u}\,f^\prime(0)-\frac{\hat{u}^2}{2}f^{\prime\prime}(0))(I_1-I_{1,0})}{\hat{u}^3} \,d\hat{u}\,,\nonumber
\end{align}
where  $I_{1,1}$ and $I_{1,2}$ indicate the first and second derivatives of $I_1$ with respect to $\hat u$ evaluated at $\hat u=0$.
If we now denote by $P_{I_1}^{(n)}$ the Taylor expansion of $I_1$ around $\hat u=0$ through order $\hat u^{n-1}$,  we see  
that the structures
\begin{equation}\label{eq:er23}
\dfrac{I_1-P_{I_1}^{(n)}}{\hat{u}^n}
\end{equation}
are regular at $\hat u=0$ for finite $\rho$ and determine the form of the resulting distributions. In analogy with what we did in  Eq.~\eqref{pippo3} they can be expressed in terms  of a divergent piece with power singularities in $\hat u$ and a residual finite (or integrable-divergent) function
\begin{equation}
\dfrac{I_1-P_{I_1}^{(n)}}{\hat{u}^n}=D_n(\hat{q}^2,\hat{u},\rho)+B_n(\hat{q}^2,\hat{u}) + O(\rho),\label{Bndef}
\end{equation}
where $D_1(\hat{q}^2,\hat{u},\rho)=\ln(\rho/(\hat u+ \rho))/\hat u w$ and $B_1(\hat{q}^2,\hat{u})=B(\hat{q}^2,\hat{u})$, following the notation of Eq.~\eqref{pippo3}.
The integrals of the divergent pieces
\begin{equation}
\mathcal{D}_n=\int_0^1 D_n(1-w,\hat{u},\rho) \,d\hat{u}
\end{equation}
converge for $\rho\neq 0$ and can be expanded in powers of $\rho$. The relevant ones are given by
\bea
\mathcal{D}_{1}&=& a=-\frac{\pi^2+3 \ln^2 \rho}{6w} + O(\rho),\\
\mathcal{D}_{2}&=& \!\!- \frac{1+\ln \rho}{w \rho}+ \frac{w-2}{2w^3} \Big(\ln^2\rho+\frac{\pi^2}3\Big)-\frac{w^2-w+2}{w^3} \ln\rho\nn\\
&&+\frac{w^2+w-4+2\ln w}{w^3}+ O(\rho),\\
\mathcal{D}_{3}&=&\!\!\frac{1\!+\!2\ln \rho}{4w \rho^2}+\frac{2(w\!-\!2) \ln \rho-4 w^2+3 w-6}{4
   w^3\rho}-\frac{w^2-6w+6}{2w^5}\Big(\!\ln^2\rho+\frac{\pi^2}3\!\Big)+
   \frac{25\!-\!2w}{w^5}\ln w
\nn\\
   &&-\frac{w^4-2w^3+7w^2-18w+18}{2w^5}\ln\rho+\frac{3w^4-21 w^2+158w-316}{12w^5}+ O(\rho).\eea
Let us now return to (\ref{1u3decomp}) and consider the last term on the rhs. We can rewrite
$I_1-I_{1,0}$ using (\ref{pippo3}) as
\be
I_1-I_{1,0} =\frac1{w} \Big( \ln \rho -\ln \hat u\Big) + \hat u \,B_1(\hat q^2, \hat u)+O(\rho)
\ee
because the rest of the integral is regular at $\hat u =0$. The first two terms correspond to plus distributions, and using also the $\rho$ expansion of $I_{1,0}$ (\ref{limitI10}) we arrive at
\bea
\int f(\hat{u})I_1\left[\dfrac{1}{\hat{u}^3}\right]_+d\hat{u}\;\!\!\!&=&\!\!\!\int_0^1 f(\hat u)\bigg[\frac{2\ln w}{w}\Big[\frac1{\hat u^3}\Big]_+\!- \frac1{w}\Big[\frac{\ln\hat u}{\hat u^3}\Big ]_+\!+\! \mathcal{D}_3\delta(\hat u)-\mathcal{D}_2\delta'(\hat u)+\frac{ \mathcal{D}_1}2\delta''(\hat u)\bigg]d \hat u\nn\\
&&+\int_0^1 \bigg[\dfrac{f(\hat{u})-f(0)-\hat{u}\,f^\prime(0)}{\hat{u}^2}\,B_1+f(0) B_3+f'(0) B_2\bigg]d\hat{u}\label{eq:Bn}
\eea
where the arguments $(\hat q^2, \hat u)$ of the $B_i$ are understood. We can then expand $B_{1}$ in powers of $\hat u$, as reported in (\ref{eq:Bexp}),
\be
B_1= B_1^{(0)}+B_1^{(1)} \hat u +...  ,
\ee
and notice that the higher orders in the $\hat u $ expansion of $B_{2,3}$ have to be related to those of $B_1$, see (\ref{Bndef}). In particular, one finds
\bea
&&a_2=B_2-\frac{B_1-B_1^{(0)}}{\hat u}=\frac{3+\ln \frac{\hat u}{w^2}}{w^3}, \\
&&a_3=B_3-\frac{B_1-B_1^{(0)} - B_1^{(1)}\hat u}{\hat u}=\frac{78-11w-3(w-12)\ln \frac{\hat u}{w^2}}{3w^5}+\frac{25+6\ln \frac{\hat u}{w^2}}{6w^5}\,\hat u, \nn
\eea
so that the second line of (\ref{eq:Bn}) becomes
\bea
&&\int_0^1 \bigg[f(\hat u) \bigg( B_1^{(0,c)}\Big[\frac1{\hat u^2}\Big]_+
+B_1^{(0,l)}\Big[\frac{\ln \hat u}{\hat u^2}\Big]_+
+ B_1^{(1,c)} \Big[\frac1{\hat u}\Big]_++B_1^{(1,l)}\Big[\frac{\ln \hat u}{\hat u}\Big]_+\nn
\\
&&\qquad\qquad\qquad+\frac{B_1-B_1^{(0)}-B_1^{(1)} \hat u}{\hat u^2}
\bigg)+a_2 \, f'(0) + a_3\, f(0)\bigg]d\hat{u},\label{eq:last1u3}
\eea
where $B_1^{(n,l)}$ is the coefficient of $\ln\hat u$ in $B_1^{(n)}$, and $B_1^{(n,c)}$ its remainder: $B_1^{(n)}= B_1^{(n,l)}\ln\hat u+ B_1^{(n,c)}$.
Combining Eqs.~\eqref{eq:Bn} and \eqref{eq:last1u3} we see that in the massless limit $I_1 [\frac1{\hat u^3}]_+$ can be expressed in terms of various distributions, with coefficients
that contain divergences as strong as $1/\rho^2$. We recall that similar lower order plus distributions can be reduced using (for $n\ge1$)
\be
\hat u \Big[\frac1{\hat u^n}\Big]_+= \Big[\frac1{\hat u^{n-1}}\Big]_+\, . \label{plusid}
\ee
It is also  worth noting that the coefficients $d_i, e_i, f_i$ in \eqref{Wipi1} contain inverse powers of
$\hat u + \rho$, which may generate additional divergences. However, combining algebraic manipulations like
\be
\frac{\rho}{(\hat u +\rho)^4}=\frac{1}{(\hat u +\rho)^3}-\frac{\hat u}{(\hat u +\rho)^4}
\ee
with  \eqref{plusid}, one can remove any such inverse power from the coefficients of the plus distributions.

We are finally ready to take the massless limit for all the terms in \eqref{Wipi1}.
As expected all power and logarithmic divergences in $\rho$ cancel out in the form factors $W_i^{(\pi,1)}$. The final results are given in the Appendix.

For what concerns the $O(\as)$ corrections to the coefficients of the chromomagnetic
matrix element, namely $W_i^{(G,1)}$, they can be computed from the results of
Ref.~\cite{Alberti:2013kxa} using the same procedure we have followed for $W_i^{(\pi, 1)}$.
The results are also given in the Appendix.


\section{Reparametrization Invariance relations}
Reparametrization Invariance (RI) \cite{RPI,Manohar:2000dt}
connects different orders in the heavy quark expansion.
This in general implies relations among the coefficients of a number of operators, see {\it e.g.}\ \cite{Fael:2018vsp},
but we are interested only in the way RI links the coefficient of the kinetic operator to the coefficient of the leading, dimension 3 operator.  In the total rate this corresponds to a rescaling factor
$1-\mu_\pi^2/2 m_b^2$ on the leading power result,
which corresponds to the relativistic dilation factor of the lifetime of a moving quark and applies at any order in perturbation theory.
The relations for differential distributions have been studied by
 Manohar who has derived  RI relations \cite{Manohar:2010sf}
directly at the  level of
the structure functions $W_i$. They are  valid to all orders in perturbation theory and give the
 coefficient of  the $O(\as \mu_\pi^2/m_b^2)$ corrections  in terms of the  $O(\as)$ coefficient and its derivatives:
\bea
W_1^{(\pi,1)}&=&- W_1^{(n)} +\frac23 W_2^{(1)}-2\hat q_0\, \frac{d\,W_1^{(1)}}{d\,\hat u}+\frac{\lambda}3  \frac{d^2 W_1^{(1)}}{d\,\hat u^2},\nn\\
W_2^{(\pi,1)}&=&\frac53 W_2^{(1)} -\frac{14}3\hat q_0\, \frac{d\,W_2^{(1)}}{d\,\hat u} +\frac{\lambda}3  \frac{d^2 W_2^{(1)}}{d\,\hat u^2},\label{RPI}\\
W_3^{(\pi,1)}&=& -\frac{10}3\hat q_0\, \frac{d\,W_3^{(1)}}{d\,\hat u}+\frac{\lambda}3  \frac{d^2 W_3^{(1)}}{d\,\hat u^2}.\nn
\eea
These relations have been verified in \cite{Alberti:2012dn}
for decays to charm. Here we verify them in the massless case as well. To this purpose
we need  the first two derivatives of the plus distributions in Eq.~(\ref{NLO}).
They can be re-expressed in terms of the higher order plus distributions
introduced in Eq.~(\ref{plusn}) and of delta functions:
\bea
\left[\frac1{\hat u}\right]'_+&=& - \left[\frac1{\hat u^2}\right]_+ +\delta(\hat u) - \delta'(\hat u),\\
\left[\frac1{\hat u}\right]''_+ &=& 2 \left[\frac1{\hat u^3}\right]_+
-\delta(\hat u)+2\,\delta'(\hat u) -\frac32  \delta''(\hat u),\\
\left[\frac{\ln \hat u}{\hat u}\right]'_+&=&  \left[\frac1{\hat u^2}\right]_+ -\left[\frac{\ln \hat u}{\hat u^2}\right]_+,\\
\left[\frac{\ln \hat u}{\hat u}\right]''_+&=& -3\left[\frac1{\hat u^3}\right]_+ + 2 \left[\frac{\ln \hat u}{\hat u^3}\right]_+ + \delta(\hat{u}) - \delta^\prime(\hat{u}) + \dfrac{1}{2}\delta^{\prime\prime}(\hat{u}),\label{eq:derivatives}
\eea
where we have neglected terms that do not contribute upon integration in the physical range
(\ref{physrange}). The coefficients $W_i^{(\pi,1)}$ obtained from Eq.~(\ref{NLO})
using the
RI relations agree with the results given in the Appendix.
On the other hand, the coefficients $W_i^{(G,1)}$ cannot be derived from RI relations.

\section{Applications}

\begin{figure}[t]
\begin{center}
\includegraphics[width=0.475\textwidth]{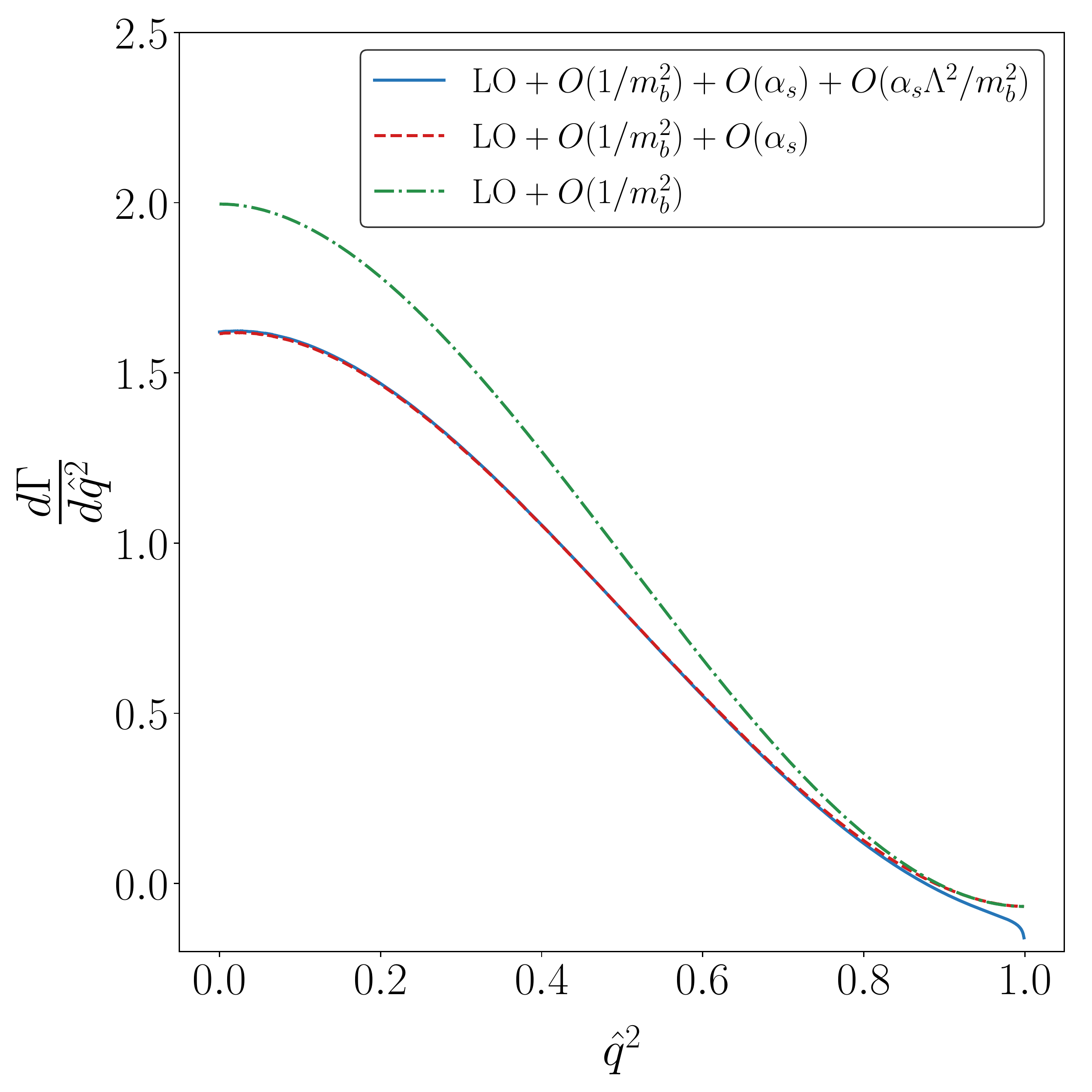}
\hspace{5mm}
\includegraphics[width=0.475\textwidth]{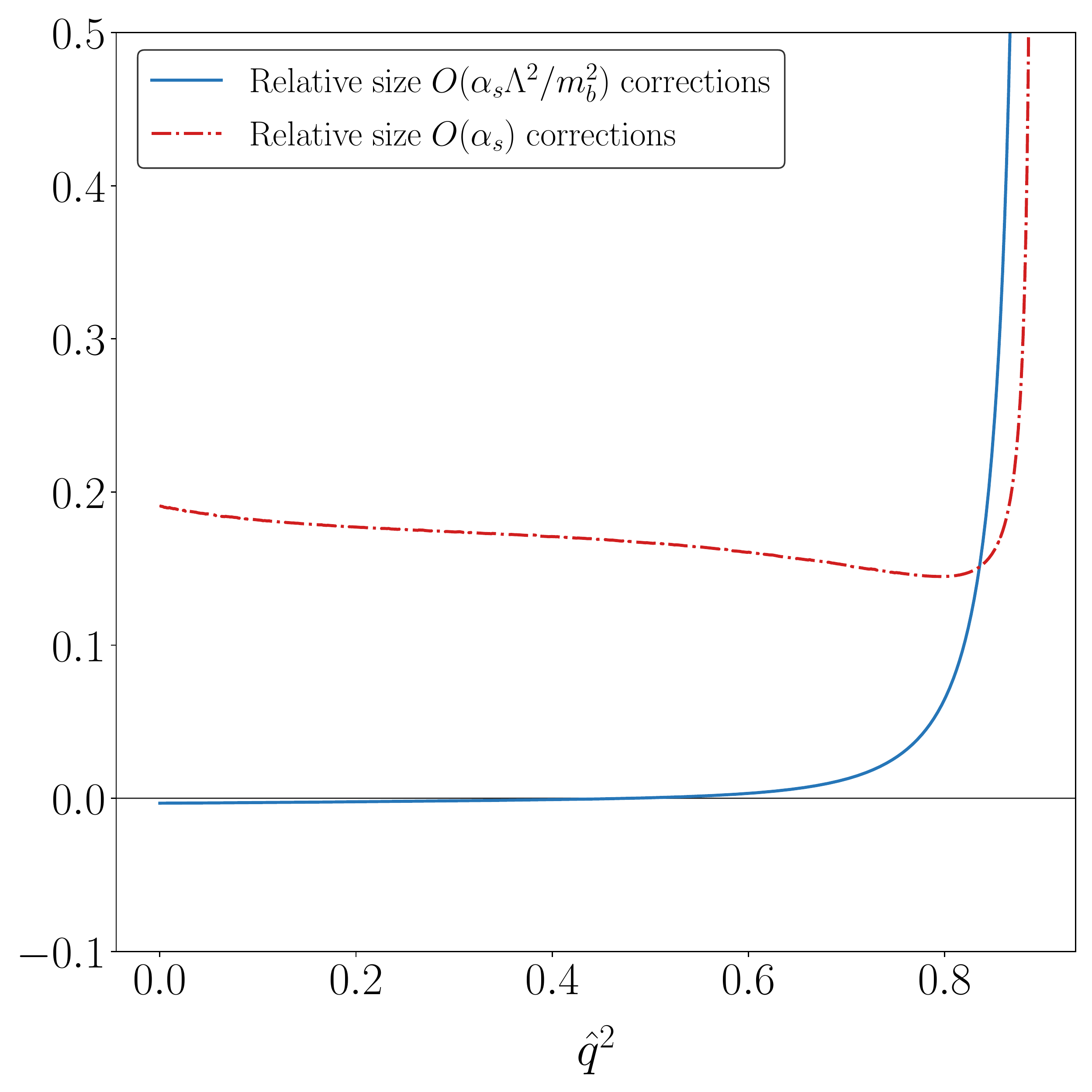}
\end{center}
\caption{Left panel: $\hat q^2$ distribution in units $\Gamma_0$ at tree level (dashed-dotted line), including  $O(\alpha_s)$ corrections (dashed line) and including also $O(\alpha_s\Lambda^2/m_b^2)$ corrections (solid line). Right panel: relative size of the $O(\alpha_s)$ (dashed-dotted line) and $O(\alpha_s\Lambda^2/m_b^2)$ (solid line) corrections.}
\label{fig:q2}
\end{figure}

The results for $W_i^{(\pi,1)}$ and $W_i^{(G,1)}$ in the massless case can be employed in Eq.~(\ref{rate}) to compute the $O(\as\Lambda^2/m_b^2)$ corrections to the total rate  and to the moments of various differential distributions in $B\to X_u \ell\nu$. We first compute the total rate in the pole mass scheme and find
\be
\Gamma(B\to X_u\ell \nu)=\Gamma_0\Big[
\Big(1-2.41 \frac{\as}{\pi}\Big) \Big(1-\frac{\mu_\pi^2}{2m_b^2}\Big) -\Big(\frac32+4.98 \frac{\as}{\pi}\Big)\frac{\mu_G^2(m_b)}{m_b^2}
\Big],\label{eq:width}
\ee
where $\Gamma_0=G_F^2 |V_{ub}|^2 m_b^5/192\pi^3$ is the lowest order result, and the $O(\as)$ contributions are a standard result, see \cite{DeFazio:1999ptt}.
As already discussed, the $O(\as \mu_\pi^2/m_b^2)$ corrections are dictated by RI.
The non-trivial $O(\as \mu_G^2/m_b^2)$ correction to the total width is sizeable and amounts
to almost a quarter of the $O(\mu_G^2/m_b^2)$ correction, but comes
with a sign opposite to the $O(\as \mu_\pi^2/m_b^2)$ correction and tends to cancel it. Using $\as=0.22$, $m_b=4.55$GeV, $\mu_\pi^2=0.43$GeV$^2$ and $\mu_G^2(m_b)=0.35$GeV$^2$,  the total shift induced by  $O(\as\Lambda^2/m_b^2)$ contributions amounts to -0.4\%.
Our result for the $O(\as \mu_G^2/m_b^2)$ correction to the total width agrees with
Ref.~\cite{Mannel:2015jka}, where the $O(\as \mu_G^2/m_b^2)$ correction to the total width and to a few $q^2$ moments has been computed in an expansion in $m_c/m_b$, and the limit $m_c\to 0$ can be read from the first term in the expansion.

We have also computed the $\hat{q}^2$ distribution. It is displayed in Fig.~\ref{fig:q2}, using the same inputs as above. One observes that the total correction is very small over the whole $\hat{q}^2$ range, except close to the endpoint, which is a region dominated by soft dynamics.

As explained in the Introduction, the rate subject to experimental cuts  is  determined by shape functions (SFs) that satisfy   OPE constraints.  Indeed,  the corrections we have computed in this paper have an important effect on these constraints, which are related to the $\hat{q}_0$-moments of the form factors $W_i$. In the GGOU framework of Ref.~\cite{Gambino:2007rp}, a $q^2$-dependent SF is associated to each form factor $W_i$, which is in turn described by the convolution formula
\begin{equation}\label{eq:er30}
W_i(\hat q_0,\hat q^2,\hat \mu)=
\int F_i(\kappa,\hat q^2,\hat \mu)W_i^{pert}\left[\hat q_0-\frac{\kappa}{2}\left(1-\frac{\hat q^2 m_b}{ m_B}\right),\hat q^2,\hat\mu\right]d\kappa.
\end{equation}
Here $W_i^{pert}$ represents the purely perturbative part of the structure functions in the kinetic scheme, and the structure function $W_i$ depends on a hard cutoff $\mu= \hat \mu m_b\sim 1$GeV that is meant to separate perturbative and non-perturbative contributions. While the SFs $F_i$ describe all nonperturbative physics, the $\hat q_0$-moments (or equivalently $\uh$-moments) of (\ref{eq:er30}) must match their OPE prediction, which can be shown to place constraints on the SFs moments, $\int \kappa^n F_i(\kappa, \hat q^2, \hat \mu) d\kappa$.  This matching has been performed at  the tree-level in \cite{Gambino:2007rp} but the $O(\as \Lambda^2/m_b^2)$ calculation of this paper permits to extend it at $O(\as)$.

In the following we compute the first three $\hat{q}_0$-moments up to $O(\as\Lambda^2/m_b^2)$ for fixed $\hat{q}^2$, leaving a detailed discussion of the constraints on the SFs to a future publication, which will also deal with the phenomenological consequences.

Let us consider the central moments of the power suppressed contributions
\begin{align}
J_{i,X}^{(n,j)}(\hat q^2)&= \int_{0}^{\infty} (\hat q_0-\hat q_0^{\,max})^n  \,W_i^{(X,j)} (\hat{q}_0,\hat{q}^2)\,d \hat q_0, \label{q0momPow}
\end{align}
where $j=0,1$ and $X=\pi, G$. While the upper endpoint in the real radiation contributions is %
\begin{equation}
\hat q_0^{\,max}=\frac{1+\hat q^2}{2},
\end{equation}
and follows from the $\theta(\hat u)$ in the expressions for $W_i^{(X,j)}$,
the lower boundary for the integrals in Eq.~(\ref{q0momPow}) is an arbitrary choice, which coincides with the physical range of the semileptonic $B$ decay
\be
\sqrt{\hat q^2}\le \hat q_0\le  \frac{1+\hat q^2}{2} \label{q0range}
\ee
only at $\hat q^2=0$. On the other hand, we note that the physical range (\ref{q0range}) becomes narrower for larger $\hat q^2$ and vanishes at the maximal value, $\hat q^2=1$.
In order to include in the integration most of the nonperturbative part of the spectral function,  we therefore consider a larger range.\footnote{The $\hat u$ range corresponding to the integrals  in Eqs.~(\ref{q0momPow}) is $(0, 1+\hat q^2)$, which extends beyond the range of the plus distributions defined in (\ref{plusn}). This implies that a redefinition based on Eq.~(3.4) of \cite{Alberti:2012dn} is necessary; it takes a form analogous to that shown in Eq.~(2.18) of that paper. }
This will be important for placing meaningful constraints on the SFs in the GGOU framework \cite{Gambino:2007rp}, where there is a $q^2$-dependent SF associated to  each form factor $W_i$, and  the $J_{i,X}^{(n,i)}$ are the building blocks necessary to achieve that.
\begin{figure}[t]
\begin{center}
\includegraphics[width=0.47\textwidth]{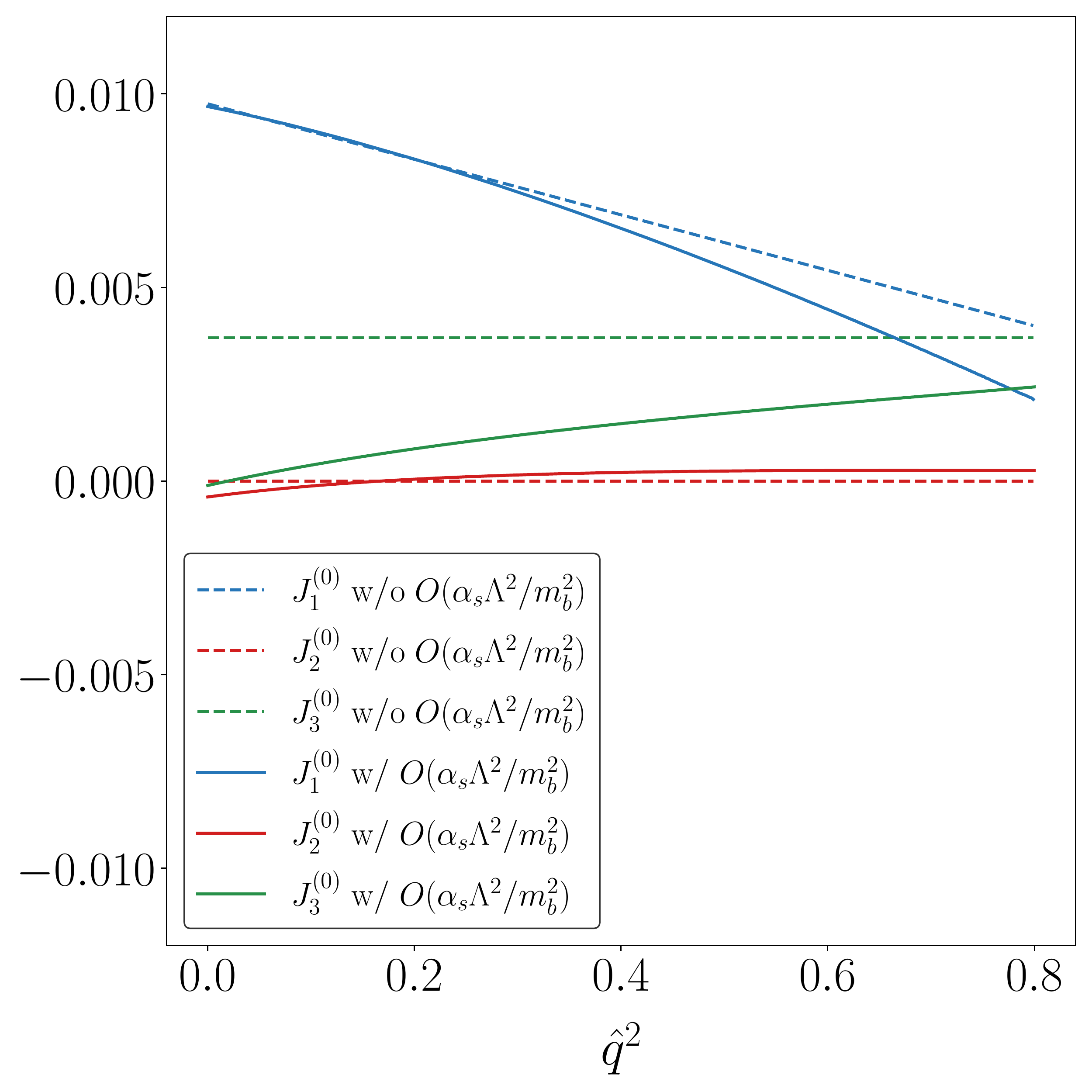}\hspace{5mm}
\includegraphics[width=0.47\textwidth]{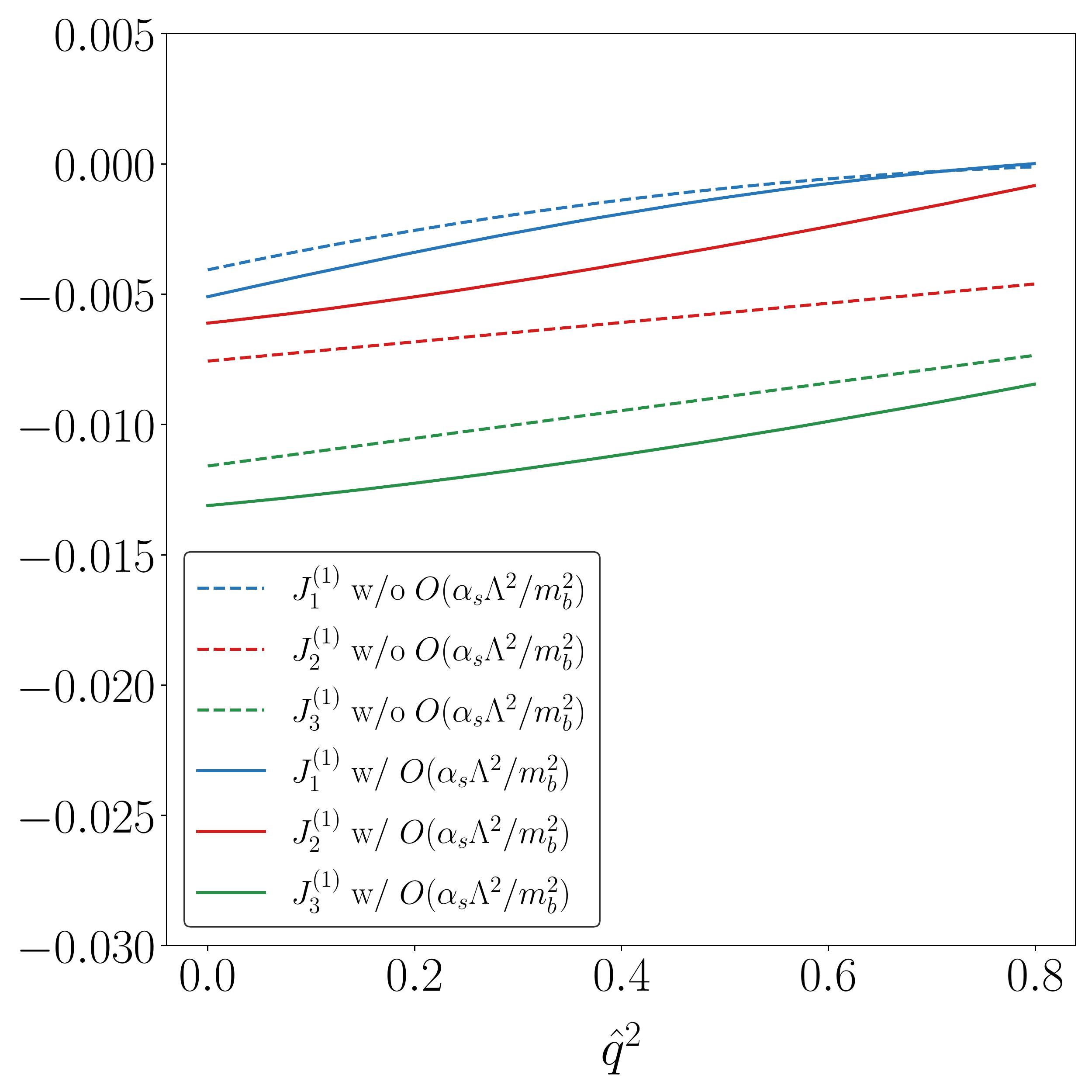}
\includegraphics[width=0.47\textwidth]{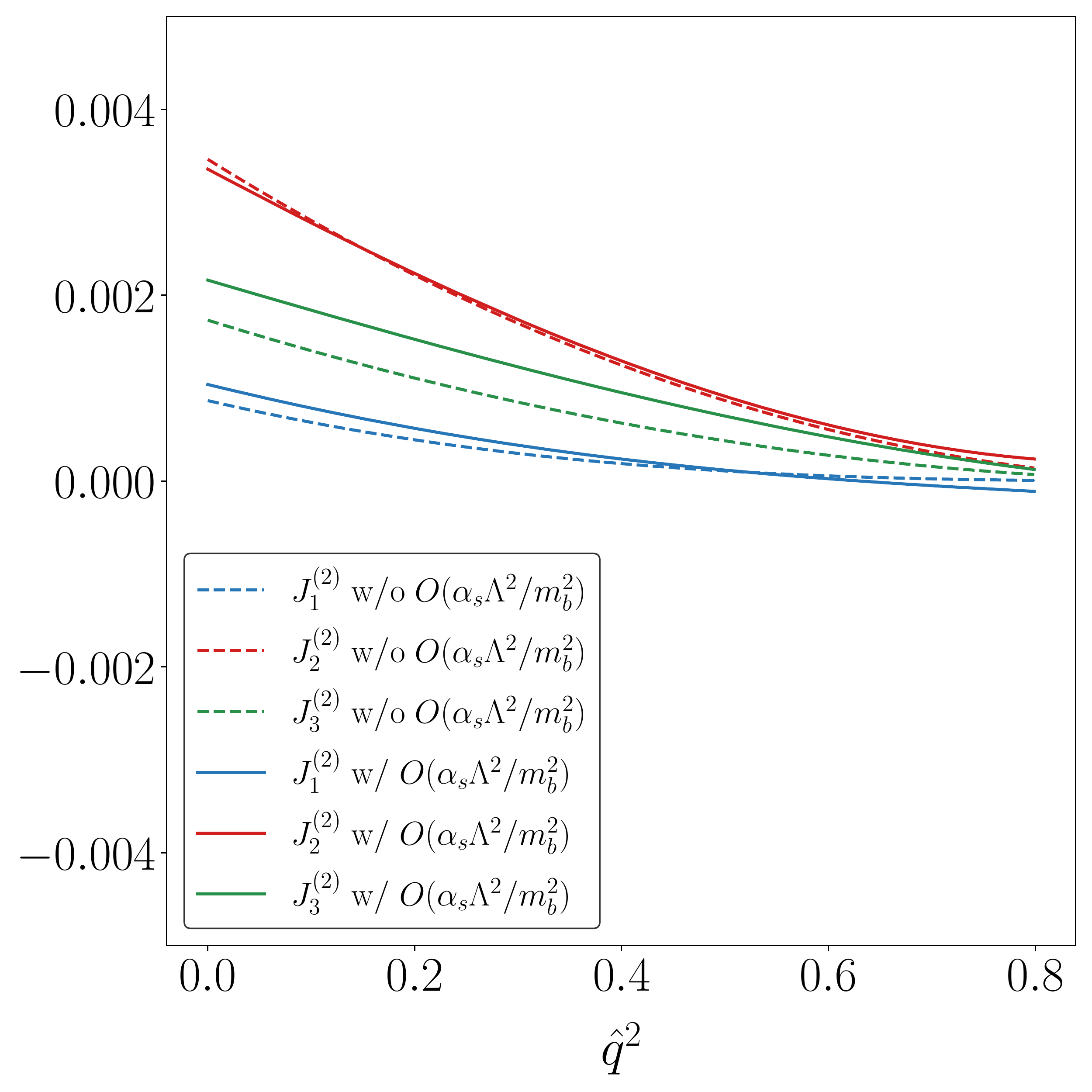}
\end{center}
\caption{Power corrections to  the lowest, first and second $\hat q_0$-moments of the form factors $W_i(\hat q_0, \hat{q}^2)$ on $\hat{q}^2$, with (solid lines) and without (dashed lines) the $O(\alpha_s\Lambda^2/m_b^2)$ corrections.}
\label{fig:Jn}
\end{figure}

The tree-level expressions  $J_{i, X}^{(n,0)}(\hat q^2)$ are given in the Appendix of Ref.~\cite{Gambino:2007rp}, while the $O(\as)$ and $O(\as\Lambda^2/m_b^2)$ corrections can be computed from the expressions for $W_i^{(1)}$ and $W_i^{(X,1)}$, respectively.  In the Appendix we provide analytic results for 
$J_{i,\pi}^{(n,1)}(0)$ and $J_{i,G}^{(n,1)}(0)$.
Let us also introduce
\begin{align}
J_{i}^{(n)}(\hat q^2)&= 
\dfrac{\mu_\pi^2}{2m_b^2}J_{i,\pi}^{(n,0)}(\hat q^2)+\dfrac{\mu_G^2}{2m_b^2}J_{i,G}^{(n,0)}(\hat q^2)
+\dfrac{\alpha_s}{\pi}\left[
\dfrac{\mu_\pi^2}{2m_b^2}C_F J_{i,\pi}^{(n,1)}(\hat q^2)+\dfrac{\mu_G^2}{2m_b^2}J_{i,G}^{(n,1)}(\hat q^2)\right].\nonumber
\end{align}
In Fig.~\ref{fig:Jn} we compare the moments $J_{i}^{(n)}$
with and without the $O(\as\Lambda^2/m_b^2)$ corrections. We employ again the same inputs as before.
The $O(\as\Lambda^2/m_b^2)$ corrections to the zeroth moments  are relatively small in most of the $\hat q^2$ range for $J_{1,2}^{(0)}$, and significant for $J_3^{(0)}$. We observed  that if
we compute the zeroth moments in the physical range (\ref{q0range}), the impact of $O(\as\Lambda^2/m_b^2)$ corrections is much larger, with the exception of the smallest values of $\hat q^2$.
The reason why this does not imply large $O(\as\Lambda^2/m_b^2)$  corrections to
the total width and the $q^2$ spectrum has to do with the prefactors of $W_i$ in the differential width.
For what concerns the higher moments, the $O(\as\Lambda^2/m_b^2)$ corrections are generally moderate, but significant  in a few cases, as a consequence of cancellations occurring at the tree level.


\section{Summary}

We have presented an analytic calculation of the $O(\as)$ corrections to the Wilson
coefficient of the kinetic and chromomagnetic operators in inclusive semileptonic decays without charm. Our results agree with reparametrization invariance relations and with a previous result on the total width. We find small corrections to the total rate and to the $q^2$ spectrum, generally below 1\% and more significant corrections to some of the moments of the form factors.
Our results place constraints on the SFs that describe   $B\to X_u \ell\nu$ decays, and in particular allow for a determination of the perturbative corrections to
their moments. This may prove useful in view of the higher precision expected
at Belle II.

\section*{Acknowledgements}
We are grateful to Leonardo Vernazza for useful correspondence and to Enrico Lunghi for pointing out to us a typographical error in Eq.~\eqref{eq:derivatives}, that we corrected in the current version of the paper. BC and PG are  supported in part by the Italian Ministry of University and Research (MUR) under grant PRIN 20172LNEEZ. The work of SN is supported by the Science and Engineering Research Board, Govt. of India, under the grant CRG/2018/001260.

\section*{Appendix}
\setcounter{equation}{0}
\renewcommand{\theequation}{A.\arabic{equation}}
\renewcommand{\thesubsection}{\Alph{subsection}}
\setcounter{section}{0}

In this Appendix we report the main results of  our calculation. In particular, the perturbative corrections to the power corrections related to the kinetic operator are given by
\bea
W_1^{(\pi,1)}&\!\!\!=& \!\!w \Big[B_{1} -\frac{C}2+
\frac{5w-2}{12} \Big[\frac{1}{\hat u^2}\Big]_+ + \Big(\frac{16+3w-10 w^2}{12}-\frac{8 w^3-w^2-14 w+8 }{6 (1-w)}\ln
   w\Big)\delta'(\hat u)\Big] \nn\\
&&
\!\!\!-\frac43(2\!-\!w) \Big(\Big[\frac{\ln \hat u}{\hat u}\Big]_+\!\!+ L_w\delta(\hat u)\Big) +\Big(\frac{8}3(2\!-\!w) \ln w-\frac{4+\!18w\!-\!13w^2}{6w}\Big)\Big[\frac{1}{\hat u}\Big]_+\!\!+{\cal R}_1^{(\pi)} \theta(\hat u) \nn\\
&& \!\!\!
+\Big( 
\frac{13 w}{12}-\frac16 - \frac1{3 w}  - \frac{w^2}{12} + 	\frac{w^3}4
+\frac{4 + 6 w - 13 w^2 + 3 w^3 + 2 w^5}{3w(1-w)}\ln w
\Big) \delta(\hat u)
\eea
\bea
W_2^{(\pi,1)}\!\!&=&\!4 B_{2} +6C +\frac{9w-10}3 \Big[\frac{1}{\hat u^2}\Big]_+ +
\Big(
\frac{4 + 6 w + 16 w^2}3\ln w-\frac{22-21w+10w^2}3
\Big)\delta'(\hat u)\nn\\
&&
\!\!\!+
\Big(
w^2+\frac{116}{3 w^2}-7 w-\frac{50}{w}+\frac{88}{3}
-4\,\frac{42 - 34 w + 17 w^2 - 6 w^3 + 2 w^4}{3w^2}
\ln w
\Big) \delta(\hat u)\nn\\
&& \!\!\!+\Big(\frac{10}3 -\frac{68}{3w}+\frac{28}{w^2}\Big)\Big[\frac{1}{\hat u}\Big]_++ {\cal R}_2^{(\pi)} \theta(\hat u)
\eea
\bea
W_3^{(\pi,1)}&\!\!=&\!\! 2 B_{3} + C+\Big(
\frac{7w}6-1
\Big)\Big[\frac{1}{\hat u^2}\Big]_++ \Big(
\frac53 (1 - w) w + \frac{w (6 + 3 w - 8 w^2)}{3(1-w)} \ln w
\Big)\delta'(\hat u)\nn\\
&&
\!\!\!+2\Big[\frac{\ln \hat u}{\hat u}\Big]_+\!\!+\Big(
\frac{19}6-\frac2{w}+\frac4{w^2}-4\ln w
   \Big)\Big[\frac{1}{\hat u}\Big]_+  \!\! + \Big[
   2L_w+\frac{w^2}{2}+\frac{14}{3 w^2}-\frac{11 w}{6}-\frac{20}{3 w}\nn\\
&&  \!\!\! +\frac{41}{6}+\Big(\frac{7w-6}{1-w}+\frac43 w  -\frac43 w^2-\frac8{w^2}+\frac4{w}\Big)
    \ln w
   \Big] \delta(\hat u)
   + {\cal R}_3^{(\pi)} \theta(\hat u)
\eea

\bea
&B_{i}=\frac{w^2}6\bigg( \Big[\frac74  -2L_w-\frac{2-w}{1-w}\ln w + \delta_{i2} \frac{\ln w}{1-w} \Big] \delta''(\hat u)
-4\Big[\frac{\ln \hat u}{\hat u^3}\Big]_++(8\ln w-1) \Big[\frac{1}{\hat u^3}\Big]_+
\bigg),\nn\\
&C= \frac{2(2-w)}3 \bigg( -\Big[\frac{\ln \hat u}{\hat u^2}\Big]_+ +2 \ln w\Big[\frac1{\hat u^2}\Big]_++ L_w\,\delta'(\hat u) \bigg)\\
&L_w= {\rm Li}_2( 1-w)+2 \ln^2 w +\frac{\pi^2}3\nn
\eea
\bea
{\cal R}_1^{(\pi)}&\!\! \!=&\! \! \frac{  (4\hat u-w)(2-w)\hat u+2w^3}{3\hat u^3}
\ln \frac{\hat u}{w^2} +\Big[\frac{2w^6}{3\hat u^3}+\frac{7w^5}{3\hat u^2}-
\frac{14-5\hat u}{3\hat u^2}w^4  - \frac{13 \hat u+32}{6 \hat u}w^3\nn\\
&&
-\frac{23 \hat u^2-36 \hat u-48}{6 \hat u}w^2-
   (13 \hat u^2-58 \hat u+36) \frac{w}6-\frac{\hat u}{6} (3
   \hat u^2-26 \hat u+8)
\Big]\frac{{\cal I}_1}{\tilde \lambda}\\
&&-\frac{4 w^2}{3 \hat u^2}+\frac{2 \hat u^2+2 \hat u w-13 \hat u+17
   w-28}{3 \tilde\lambda}+\frac{4 w}{3
   \hat u^2}+\frac{2}{3 \hat u w}-\frac{7 \hat u+8}{12 \hat u}\nn
\eea
\bea
{\cal R}_2^{(\pi)}\!\!&\!=& \!\!\frac{  12(2\!-\!w)\hat u+8w^2}{3\hat u^3}
\ln \frac{\hat u}{w^2}
 + \bigg[ w\Big(\frac{8w^4}{3\hat u^3}-\!\frac{40}3-\!\frac{14w}3
\!-\!2\hat u +\frac{(4w\!-\!8)w^2}{\hat u^2}-4\frac{8-8w+w^2}{\hat u}\Big)\nn\\
&&+68+\!60\hat u-\!\frac4{\tilde\lambda}\big(15\hat u^3-35\hat u^2 -76\hat u +14w+63 w \hat u +19 w \hat u^2\big)\bigg]
\frac{{\cal I}_1}{\tilde \lambda}+\frac{16(1+2\hat u)}{3\hat u^2} \nn\\
&&  -\frac{16 w}{3 \hat u^2}-\frac{28}{\hat u w^2}+\frac{68}{3 \hat u w}
-\frac{2 \left(9 \hat u^2+50 \hat u w-201 \hat u+86 w-78\right)}{3
   \hat u \tilde\lambda}\\
   &&-\frac{4 \left(2 \hat u^2 w+2 \hat u^3+11 \hat u^2+49 \hat u w-81
   \hat u+45 w-28\right)}{\tilde\lambda^2}\nn
 \eea
 \bea
{\cal R}_3^{(\pi)}&\!\! \!=&\!-\frac{2(3\hat u^2-(2-w) \hat u -2w^2)}{3\hat u^3} \ln \frac{\hat u}{w^2}+
\Big(8 w-\frac{13}3 w^2-4 -\frac{10}3 \hat u (w-2)-\hat u^2\nn\\
&&+
\frac{10 \hat u (w-2) w^3+4 w^5}{3 \hat u^3}\Big) \frac{{\cal I}_1}{\tilde \lambda}
-\frac{2 \left(7 \hat u^2+11 \hat u w-19 \hat u+17 w-16\right)}{3
   \hat u \tilde\lambda}
   \\
    &&
   -\frac{8 w}{3
   \hat u^2}+\frac{8 (\hat u+1)}{3 \hat u^2}-\frac{4}{\hat u
   w^2}+\frac{2}{\hat u w}\nn
\eea
In the above expressions the coefficients of the derivatives of $\delta(\hat u)$ have been
reduced using integration by parts identities like
\be
f(\hat u) \,\delta '' (\hat u)= f(0)\, \delta''(\hat u) -2 f'(0)\, \delta'(\hat u) + f''(0)\, \delta(\hat u),
\ee
as well as  identities such as \eqref{idplus} and (\ref{plusid}).

The analogous results for the coefficients of the matrix element of the chromomagnetic operator are
\bea
W_1^{(G,1)}&\!\!\!=& \!\!-\frac23 w \bigg[G_{1}  + \Big(
 \mbox{\small $\frac{C_F}{4}$}  \Big(1+8w-5\frac{w^2\ln w}{1-w}\Big)-  \mbox{\small $\frac{C_A}{4}$} (1+2w)
\Big)\delta'(\hat u)\\
&&
\!\!+\mbox{\small $C_F(5+\frac2{w^2}-\frac2{w})$} \Big(\Big[\frac{\ln \hat u}{\hat u}\Big]_+\!\!-\!2\ln w\Big[\frac{1}{\hat u}\Big]_+\!\!+\! L_w\delta(\hat u)\Big)\Big] \nn\\
&&\!\!-\frac23\Big( \mbox{\small $\frac{C_A}4$} (8\!-\!5w) + C_F \big(\frac{4}{w} -3+\frac{5w}4\big)\Big) \Big[\frac{1}{\hat u}\Big]_+\nn\\
&& \!\!
-\frac13\big(\mbox{\small $C_A
\frac{ 5 w^3-34 w^2+51 w-20}{2 (w-1)^2}+C_F\frac{
   10 w^5-21 w^4+7 w^3-10 w^2+28 w-16}{(w-1)^2 w}$}
\big) \ln w \, \delta(\hat u)\nn\\
&& \!\!-\frac13\big(\mbox{\small $C_A
\frac{ 2 w^4+2 w^3-3 w^2+5 w-4}{2 (1-w) w}+C_F\frac{
   35 w^3-25 w^2-10 w-8}{4 (1-w)}$}
\big)\delta(\hat u)+{\cal R}_1^{(G)} \theta(\hat u) \nn
\eea
\bea
W_2^{(G,1)}&\!\!=& \!\!-\frac83 \Big[G_{2}  + \Big(
\mbox{\small $C_F$} \Big(\frac1{w}-\frac{11}4+2w-(1-\frac{5w}4)\ln w\Big) +\mbox{\small $\frac{C_A}4$} (3-2w) \Big)\delta'(\hat u) \nn\\
&&
\!\!\!\!+\mbox{\small $2C_F\frac{w^2-w-1}{w^3}$} \Big(\Big[\frac{\ln \hat u}{\hat u}\Big]_+\!\!-\!2\ln w\Big[\frac{1}{\hat u}\Big]_+\!\!+\! L_w\delta(\hat u)\Big) \nn\\
&&\!\!\!\!+
\!\big( \mbox{\small $C_A(\frac8{w}-\frac9{2w^2}-\frac2{w^3}-\frac94)$}
 + \mbox{\small $C_F (\frac{7}{w^2} -\frac6{w^3}-\frac{5}2)$}\big) \Big[\frac{1}{\hat u}\Big]_+ \\
&&
\!\!\!\!+\big(\mbox{\small $
C_A\frac{9 w^3-56 w^2+40 w+16}{4 w^3}
  +C_F  \frac{5 w^4-6 w^3+3 w^2-12 w+12}{w^3} $}
\big)\ln w \,  \delta(\hat u)\nn\\
&&
\!\!\!\!-\big(\mbox{\small $C_A\frac{2 w^3+4 w^2-23 w+16}{4 w^2}+C_F\frac{
   35 w^4-98 w^3+134 w^2-120 w+32}{8 w^3}$}\big) \delta(\hat u) \Big] +{\cal R}_2^{(G)} \theta(\hat u)\nn
\eea

\bea
W_3^{(G,1)}&\!\!\!\!=& \!\!-\frac43 \Big[G_{3}  + \Big(
\mbox{\small $C_F$} \Big(\frac1{4}+\frac{5w}2-\frac{5w^2 \ln w}{4(1-w)}\Big) -\mbox{\small $\frac{C_A}4$} (1+w) \Big)\delta'(\hat u)\Big] \nn\\
&&
\!\!-\frac23\mbox{\small $C_F\frac{5w^2+4w+4}{w^2}$} \Big(\Big[\frac{\ln \hat u}{\hat u}\Big]_+\!\!-\!2\ln w\Big[\frac{1}{\hat u}\Big]_+\!+ L_w\delta(\hat u)\Big) \nn\\
&&\!\!-\frac23
\big( \mbox{\small $C_A(\frac2{w^2}+\frac5{w}-\frac72)$}
 + \mbox{\small $C_F (\frac{4}{w^2} -\frac{5}4)$}\big) \Big[\frac{1}{\hat u}\Big]_+ \\
&&
\!\!-\frac13\big(\mbox{\small $C_A
\frac{7 w^4-40 w^3+49 w^2-6 w-8}{ (w-1)^2
   w^2}+C_F\frac{(20 w^5-37 w^4-w^3+6 w^2+24 w-16)}{ (w-1)^2
   w^2}
 $}
\big)\ln w \,  \delta(\hat u)\nn\\
&&\!\!-\frac23
\big(\mbox{\small $C_A
   \frac{w^2-w+1}{1-w}+C_F\frac{35 w^4-85
   w^3+66 w^2-8 w-16}{4 (1-w) w^2}
   $}
      \big) \delta(\hat u)  +{\cal R}_3^{(G)} \theta(\hat u)\nn
\eea

\bea
G_i&=&\!\! \Big(1+\frac52 w-4 \delta_{i2}\Big)\bigg[\mbox{\small $C_F$}\Big(\frac{3-8\ln w}4 \Big[\frac{1}{\hat u^2}\Big]_+ + \Big[\frac{\ln \hat u}{\hat u^2}\Big]_+ - L_w \delta'(\hat u)\Big) +\mbox{\normalsize $\frac{C_A}2$}
\ln \frac{\mu}{m_b}  \delta'(\hat u)
\bigg] \nn\\
&&\!\!\!\! +C_A \bigg[\frac{1\!+\!w}2\Big[\Big[\frac{1}{\hat u^2}\Big]_+\!\!
+ \ln w\, \delta'(\hat u) \Big] \!-\!\delta_{i2}\Big(
\frac{1\!+\!2w}{2w}\Big[\frac{1}{\hat u^2}\Big]_+\!\!+\!\frac{\ln w}{w}\delta'(\hat u)
\Big)\bigg]  \!
-\! \mbox{\normalsize $\frac{3C_A}4$} \frac{w_i^{(G,0)}}{w_i^{(0)}}\ln \frac{\mu}{m_b}  \delta(\hat u)\nn\\
&&\! \!\!\!  + C_A  \Big(\frac{1+4w}{2w^2} -\frac{1+2w}{w^3} \delta_{i2}\Big) \bigg[\Big[\frac{\ln \hat u}{\hat u}\Big]_+ -2 \ln w \Big[\frac{1}{\hat u}\Big]_++ L_w \delta(\hat u)\bigg]
\eea
where
\be
w_1^{(G,0)}=-\frac{2}3(4-5w), \qquad w_2^{(G,0)}=0,\qquad w_3^{(G,0)}=\frac{10}3,
\ee
and
\bea
{\cal R}_1^{(G)}\!\!&=&\!\!\frac{C_A}3\Big[\mbox{$\frac12+\frac{\hat u +13 w-16}{\tilde\lambda}+
\frac{4w+1}{\hat u w}\ln \frac{\hat u}{w^2}$}+\Big(\mbox{$\frac{4w+1-6\hat u}{\hat u}  +2\frac{3\hat u (\hat u-3+w)+4w}{\tilde\lambda}$}\Big) {\cal I}_1\Big]\\
&&\!\!\!+\frac{C_F}3\Big[\mbox{$\frac{15 \hat u-5 \hat u w-5 \hat u^2-11  w+20 }{\tilde\lambda } -\frac{4w}{\hat u\tilde\lambda}
 -\frac{10 w}{\hat u}+\frac{8}{\hat u w}+\frac{11
   \hat u+24}{4 \hat u}$}
   +\Big(\mbox{$\frac{5 w^2}{\hat u^2}+\frac{2 (5 \hat u+1)
   w}{\hat u^2}
   +\frac{4-4w}{\hat u w}$}\Big)\ln \frac{\hat u}{w^2}\nn\\
   &&\!\!\!+\Big(\mbox{$
   \frac{8-3 \hat u^2-13 \hat u w+10 \hat u-12 w}{\tilde\lambda}+\frac{5
   w^3}{\hat u^2}+\frac{(15 \hat u+2) w^2}{\hat u^2}+\frac{3 (5
   \hat u-8) w}{2 \hat u}+\frac{5\hat u}{2} -2 $}\Big){\cal I}_1\Big]\nn
\eea
\bea
{\cal R}_2^{(G)}&\!\!\!=&\!4C_A\Big[\mbox{$\frac{16-13 \hat u^2-25 \hat u w+51 \hat u-29
   w}{\tilde\lambda^2}
     +\frac{22-15 \hat u w-9 \hat u^2+112 \hat u-32
    w }{6 \tilde\lambda \hat u}
   +\frac{w}{3 \tilde\lambda \hat u^2}
   +\frac{16
   \hat u-1}{3 \hat u^2 w}-\frac{3}{\hat u w^2}-\frac{4}{3 \hat u
   w^3}$}\nn\\
   &&\!\!\!+\mbox{$\frac{4w^2-3w-2}{3w^3\hat u}$}\ln \frac{\hat u}{w^2}+
   \Big( \mbox{$  \frac{14 \hat u^2-26 \hat u w+58
   \hat u-3 w-2}{3 \tilde\lambda \hat u}-\frac{2 \left(3 \hat u^2 w+3 \hat u^3-5 \hat u^2+20 \hat u w-25
   \hat u\right)}{\tilde\lambda^2}  -\frac{8w}{\tilde\lambda^2}+\frac{4}{3 \hat u}$}\Big) {\cal I}_1\Big]\nn\\&&\!\!\!
   +4C_F\Big[\mbox{$
   \frac{5 \hat u^2 w+42 \hat u w+5 \hat u^3-4 \hat u^2-55
   \hat u+39  w-36 }{\tilde\lambda^2 }
   +\frac{4w}{\tilde\lambda^2 \hat u}
   +\frac{53 \hat u w-20\hat u^2-155 \hat u+44 w-52}{6 \tilde\lambda
   \hat u}+\!\frac{14}{3 \hat u w^2}\!-\!\frac{4}{\hat u w^3}\!-\!\frac{10}{3
   \hat u}$}\nn\\ &&\!\!\!
      + \mbox{$\frac{4 \hat u \left(w^2-w-1\right)+(5 w-6) w^3}{3 \hat u^2 w^3} $} \ln \frac{\hat u}{w^2} + \Big(\mbox{$\frac{23 \hat u^2 w+13 \hat u^3-37 \hat u^2+47 \hat u w-58
   \hat u+20 w-8}{\tilde\lambda^2}$}
   \\&&\!\!\!
   +\mbox{$ \frac{25 \hat u^2 w+15 \hat u^3-114
   \hat u^2+76 \hat u w-150 \hat u+16 w-8}{6 \tilde\lambda
   \hat u}+\frac{5 w^2}{3 \hat u^2}+\frac{(5 \hat u-6) w}{3
   \hat u^2}-\frac{5 \hat u+8}{2 \hat u}$}\Big) {\cal I}_1\Big]\nn
\eea
\bea
{\cal R}_3^{(G)}&\!\!\!=\!\!\!&\frac{4C_A}3\Big[\mbox{$\frac{15 \hat u-3 \hat u^2-3 \hat u w-5 w-2}{2 \tilde\lambda
   \hat u}+\frac{1}{\hat u w^2}+\frac{5}{2 \hat u w}+\frac{1+4w}{2w^2\hat u}\ln \frac{\hat u}{w^2} + \frac{w-5 \hat u  - 2  w\hat u + 4 w^2}{2 \tilde\lambda \hat u} $}{\cal I}_1\Big]\quad\qquad\qquad\\&&
  +\frac{4C_F}3
   \Big[\mbox{$ \frac{2 \hat u^2+7 \hat u w-9 \hat u+3 w}{\tilde\lambda
   \hat u}+\frac{2}{\hat u w^2}-\frac{5}{\hat u} +\Big(\frac{5 w}{2 \hat u^2}+\frac{5 \hat u+2}{2 \hat u^2}+\frac{2}{\hat u
   w^2}+\frac{2}{\hat u w}\Big)\ln \frac{\hat u}{w^2}$}\nn\\
   &&  +
   \Big(\mbox{$\frac{5 \hat u^2+5 \hat u w-16 \hat u+12 w-12}{2 \tilde\lambda}+\frac{5
   w^2}{2 \hat u^2}+\frac{(5 \hat u+1) w}{\hat u^2}-\frac{5
   \hat u+8}{4 \hat u}$}\Big) {\cal I}_1\Big]\nn
\eea

The $\hat q_0$-moments of the form factors are defined in Eq.~(\ref{q0momPow}).
We first recall the tree-level results
\bea
&&J_{1}^{(0)}=\frac{1-\hat q^2}2+\frac{1+\hat q^2}3\frac{\mu_\pi^2}{m_b^2}+\frac{1-5\hat q^2}{6}\frac{\mu_G^2}{m_b^2}, \qquad J_{2}^{(0)}=2,
 \qquad J_{3}^{(0)}=1 -\frac{\mu_\pi^2}{2m_b^2}+ \frac56 \frac{\mu_G^2}{m_b^2} ,\nn\\
 &&
 J_{1}^{(1)}=\frac{1-\hat q^4}{24}\frac{\mu_\pi^2}{m_b^2}- \frac{7-12\hat q^2 +5\hat q^4}{24}\frac{\mu_G^2}{m_b^2}, \qquad
 J_{2}^{(1)}=-\frac{1+\hat q^2}2\frac{\mu_\pi^2}{m_b^2}+\frac{1+5\hat q^2}{6}\frac{\mu_G^2}{m_b^2},\nn\\
 &&
  J_{3}^{(1)}=-\frac{1+\hat q^2}{12}\frac{\mu_\pi^2}{m_b^2}+ \frac{5\hat q^2-7}{12}\frac{\mu_G^2}{m_b^2},\qquad J_1^{(2)}=\frac{(1-\hat q^2)^3}{24}\frac{\mu_\pi^2}{m_b^2},\nn\\
&&   J_2^{(2)}=\frac{(1-\hat q^2)^2}{6}\frac{\mu_\pi^2}{m_b^2}, \qquad
J_3^{(2)}=\frac{(1-\hat q^2)^2}{12}\frac{\mu_\pi^2}{m_b^2}.
\eea
Finally, here we report the  results of the  $O(\as\Lambda^2/m_b^2) $ corrections at $q^2=0$ with $\mu=m_b$:

\bea
&&J_{1,\pi}^{(0,1)}(0)=-\frac{\pi^2}{18},\qquad
J_{1,\pi}^{(1,1)}(0)=\frac{151}{48}-\frac{25\pi^2}{72},\qquad
J_{1,\pi}^{(2,1)}(0)=-\frac{191}{48}+\frac{7\pi^2}{18},\nn\\
&&
J_{2,\pi}^{(0,1)}(0)=0,\qquad
J_{2,\pi}^{(1,1)}(0)=-\frac54+\frac{\pi^2}4,\qquad
J_{2,\pi}^{(2,1)}(0)=\frac{35}{12}-\frac{11\pi^2}{36},\\
&&
J_{3,\pi}^{(0,1)}(0)=-\frac34 +\frac{\pi^2}{12},\qquad
J_{3,\pi}^{(1,1)}(0)=\frac38+\frac{\pi^2}{72},\qquad
J_{3,\pi}^{(2,1)}(0)=-\frac{11}{16}+\frac{7\pi^2}{144},\nn
\eea

\bea
&&J_{1,G}^{(0,1)}(0)=\frac{121}{18} -\frac{65\pi^2}{108},\qquad
J_{1,G}^{(1,1)}(0)=-\frac{11}{16}-\frac{13\pi^2}{216},\qquad
J_{1,G}^{(2,1)}(0)=\frac{707}{2592}+\frac{11\pi^2}{432},\nn\\
&&
J_{2,G}^{(0,1)}(0)=-\frac{59}6+\frac{25\pi^2}{27},\quad
J_{2,G}^{(1,1)}(0)=\frac{109}{36}-\frac{7\pi^2}{27},\quad
J_{2,G}^{(2,1)}(0)=-\frac{317}{72}+\frac{4\pi^2}{9},\\
&&
J_{3,G}^{(0,1)}(0)=\frac{28}3 -\frac{29\pi^2}{18},\qquad
J_{3,G}^{(1,1)}(0)=-\frac{29}9-\frac{\pi^2}{54},\qquad
J_{3,G}^{(2,1)}(0)=\frac{371}{144}-\frac{11\pi^2}{72},\nn
\eea

\end{document}